\documentclass[10pt,journal,compsoc]{IEEEtran}
\usepackage[utf8]{inputenc}
\usepackage{amsmath,amssymb,amsfonts}
\usepackage{algorithmic}
\usepackage{textcomp}
\usepackage{xcolor}

\usepackage{soul}
\usepackage{graphicx, algorithm, multirow}
\usepackage{caption}
\usepackage{cleveref}
\usepackage[tight,footnotesize]{subfigure}
\usepackage{xspace}
\usepackage{ulem}
\usepackage{booktabs}

\newcommand{\pennant}{\texttt{Pennant}\xspace}

\newcommand{\minife}{\texttt{MiniFE}\xspace}
\newcommand{\mixbench}{\texttt{MixBench}\xspace}
\newcommand{\tool}{\textsc{libnvcd}\xspace}
\newcommand{\dash}{\textsc{dashing}\xspace}
\newcommand{\matmul}{\texttt{Matmult}\xspace}
\newcommand{\baseline}{\texttt{Baseline}\xspace}
\newcommand{\bank}{\texttt{NoBankConflict}\xspace}
\newcommand{\prefetch}{\texttt{Prefetch}\xspace}
\newcommand{\unroll}{\texttt{Unroll}\xspace}

\newcommand{\compute}{\texttt{ComputeOpt}\xspace}
\newcommand{\tiling}{\texttt{Tiling}\xspace}
\newcommand{\coalescing}{\texttt{Coalescing}\xspace}

\newcommand{\score}{\texttt{score}\xspace}
\newcommand{\nvidia}{NVIDIA\xspace}
\newcommand{\libpmon}{\textsc{LibNVMON}\xspace}
\usepackage{enumitem}

\usepackage{amssymb,amsmath,amsthm}

\newcommand{\ttt}[1]{\texttt{#1}\xspace}

\usepackage{flushend}

\begin{document}
%
\title{Data-Driven Analysis to Understand GPU Hardware Resource Usage of Optimizations}%
%
%
%

\author{Tanzima~Z.~Islam,
        Aniruddha~Marathe,
        Holland~Schutte,
        Mohammad~Zaeed}

\maketitle

\IEEEtitleabstractindextext{
\begin{abstract}
With heterogeneous systems, the number of GPUs per chip increases to provide computational capabilities for solving science at a nanoscopic scale. However, low utilization for single GPUs defies the need to invest more money in expensive accelerators. Although related work develops optimizations to improve application performance, no studies have been done on how these optimizations impact hardware resource usage or average GPU utilization.
This paper takes a data-driven analysis approach to address this gap by
(1) characterizing how hardware resource usage affects device utilization, execution time, or both,
(2) presenting a multi-objective metric to identify meaningful application-device interactions that can be optimized to improve device utilization and application performance jointly,
(3) studying hardware resource usage behaviors of several optimizations for a benchmark application, and finally, (4) identifying optimization opportunities for several scientific proxy applications based on their hardware resource usage behaviors. Furthermore, we demonstrate the applicability of our methodology by applying the identified optimizations to a proxy application, which improves execution time, device utilization, and power consumption by up to 29. 6\%, 5. 3\% and 26. 5\% respectively, for several input parameters. 

\end{abstract}

}

\IEEEcompsocitemizethanks{\IEEEcompsocthanksitem Tanzima Z. Islam is with the Department of Computer Science, Texas State University, San Marcos, TX 78666.\protect\\
Email: tanzima@txstate.edu
\IEEEcompsocthanksitem Aniruddha Marathe is with the Lawrence Livermore National Laboratory, Livermore, CA 7000\protect\\
Email: marathe1@llnl.gov
\IEEEcompsocthanksitem Holland Schutte is with the Western Washington University, Bellingham, WA.\protect\\
Email: schutth@wwu.edu 
\IEEEcompsocthanksitem Mohammad Zaeed is with the Department of Computer Science, Texas State University,\protect\\
Email: cup7@txstate.edu}



%

\IEEEraisesectionheading{\section{Introduction}\label{sec:introduction}}

%
%
%
%
The design of HPC systems has been influenced by considerations of cost and power efficiency, leading to innovations in accelerator-based heterogeneous computing. For instance, according to the November 2020 Top500~\cite{top500} list, six of the top ten supercomputers make use of \nvidia GPUs (see V100 or A100). The heterogeneous computing environment necessitates the simultaneous management of shared resources such as cores, memory, power, network, and I/O while meeting specific system requirements, such as minimizing network interference, ensuring resiliency, and managing large amounts of data, making these environments inherently multi-objective. However, application scientists and performance analysts often prioritize reducing application execution time and tend to treat device utilization as a secondary metric.
Low device utilization, specifically for compute units (e.g., streaming multiprocessors or SM for GPUs), implies idle compute capacity that would otherwise be used to accomplish science. Future supercomputer design trends clearly lean towards incorporating more accelerators per node to deliver higher computational capabilities while the individual devices largely remain under-utilized. 
Hence, we argue that optimization strategies should target improving compute resource utilization and execution time. 
Typically, platform-agnostic optimizations (e.g., tiling, loop unrolling) and platform-specific ones (e.g., bank conflict resolution and memory coalescing in GPUs) in the literature~\cite{ye2009design} quantify performance improvement using only execution time. 
However, no prior work has characterized these optimizations in relation to their interactions with hardware components and their impact on execution time and compute-resource utilization, which is also referred to as a multi-objective performance metric in this paper.

All modern architectures (both CPU and GPU) include several hardware components or \textit{resources} (e.g., L1, L2, DRAM), and performance counters on each resource to describe the nature of application-device interactions, e.g., L2\_miss event counts the number of times an application did not find data in the L2 cache. If the number of L2 cache misses increases, the performance of that application will likely suffer (i.e., the execution time will also increase). The same event also indicates that the compute resources will be idle, but memory resources will be busy for data to be read from slow global device memory (high latency). On the other hand, high computational resource utilization could mean less memory traffic (but not necessarily) and improved execution time. Consequently, a trade-off space exists between compute and memory resource utilization and execution time. Understanding how optimizations influence this trade-off space is crucial for building an automated requirement-based optimization engine in the future. Such an engine, perhaps realized through an auto-tuner, can enable co-designing performance requirements from both applications. One step further towards automation is correlating the characteristics of optimizations to the trade-off space to understand \textit{why} specific optimizations help and \textit{when}. The ultimate goal of such a study would be to build a map between resource-usage characteristics and optimization. 
While prior research exists on characterizing applications' hardware usage on CPU-based systems~\cite{islam2016a,allen2016characterizing}, the same for GPUs is sparse. Performance debugging tools such as \ttt{nvprof} provide raw figures of hardware resource usage tagged with timestamps but are inadequate in providing statistical significance of the most impactful hardware resources concerning the target metric. 

\textbf{This paper fills the gap by designing metrics and a systematic methodology for characterizing application-device interactions of common optimizations using a multi-objective performance metric for GPUs.}
Explicitly focusing on improving compute resource (SM) utilization and execution time on GPUs, this paper takes a two-step approach: (1) understand how optimizations interact with an underlying platform (\textit{characterization}), and (2) identify how optimization characteristics influence a multi-objective performance metric. In this paper, we characterize six common optimizations to the \textit{matrix-matrix multiplication} (henceforth referred to as \matmul) application on GPUs, correlate the change in resource usage to optimization suggestions, identify optimization opportunities in several proxy applications using these suggestions, and finally demonstrate that these suggestions indeed improve performance and utilization--up to 29.6\% and 5.4\%, respectively. Specifically, the contributions of this paper are:
\begin{itemize}[noitemsep,nolistsep,leftmargin=0pt]
\item Design a multi-objective performance metric (Section~\ref{sec:perf-metric}) to study optimizations' impact on compute resource utilization and application performance. 
\item Build resource-usage characteristics driven optimization suggestions (Table~\ref{tab:opt-steps}) and identify optimization opportunities for several proxy applications and benchmarks (Section~\ref{sec:results}) to potentially improve application performance, compute-resource utilization, or both.
\item Apply the optimization suggestions to demonstrate the effectiveness of our approach (Section~\ref{sec:pennant}).
\end{itemize}

We applied our analysis methodology to the following Exascale Computing Project (ECP) proxy and benchmark applications--Pennant~\cite{ferenbaugh2015pennant}, MiniFE~\cite{minife}, and MixBench~\cite{mixbench}. We conducted our analysis on the \nvidia Volta platform (V100). We chose \nvidia since $2$ of the top $10$ HPC systems use \nvidia Volta GPUs. Although we focus on single-GPU characterizations and optimizations in this paper, our methodology can be applied to study a multi-GPU system. 

\section{Background}
\label{sec:background}
In this work, we leverage our publicly available performance analysis framework--\dash~\cite{islam2019toward}. This section describes the machine learning methodology for calculating the importance of hardware resource (e.g., L1, L2, DRAM) usage and application-device interactions (hardware performance events) in describing the performance of applications~\cite{islam2016a,thiagarajan2018paddle}. Since modern systems contain hundreds of performance events, inspecting even 50\% of them to infer insights into performance issues is challenging. The \dash framework takes a hierarchical approach of grouping the counters into \textit{resource groups or resources} and combining all events per resource group's contributions in reconstructing the target metric. Each resource's importance in explaining a target metric is called \textit{Resource Significance Measure (RSM)}. 
\subsection{Features, Resources, and Targets}
This work leverages on-chip registers, called hardware performance counters, to track application-device interactions (performance events or features). Performance events are used as \textit{features} because hardware bottlenecks typically constrain on-device performance. 

\textit{Resources} collectively refers to a group of performance events that describes a hardware component or a type of application-device interactions known to communicate performance issues. For example, ``L2" can be a resource where it represents a hardware component, and ``STALL" can also be a resource where it describes a phenomenon indicating performance and usage issues. 
For performance characterization, these resources define an abstract machine model such as Figure~\ref{fig:volta-arch}. The abstract machine model will differ across platforms; hence the observations made in this paper are \nvidia Volta-specific, but the methodology developed in this paper is generally applicable.

\textit{Target} describes a performance metric such as execution time, energy efficiency, and device utilization. The feature selection problem is formulated to \textit{identify features that can predict a target based on causation relationship and quantify their contributions in driving that target}. Higher contribution means a performance event described by that feature drives performance. \dash leaves the definition of features, targets, abstract machine model, and event-per-resource mapping up to the user. In Section~\ref{sec:machine-model}, we describe our contribution in creating an abstract machine model for NVIDIA Volta architecture. 

\subsection{Feature Selection using Sparse Coding}
\label{sec:rsm}
Identifying important hardware events that drive a performance metric can be formulated as a feature selection task in machine learning.
The feature selection process takes a principled approach to filter the number of events needed for diagnosing the cause of the increase in target. 
Although there is a broad spectrum of feature selection methods, the \dash framework implements an extended sparse coding technique~\cite{islam2016a} to identify important factors from high-dimensional data, which makes this approach highly interpretable
and robust to noisy data. Compared to commonly used dimensionality reduction techniques such as PCA~\cite{karamizadeh2013overview}, the output from sparse coding can be effectively used to attribute the contributions of an individual event.

The performance profile of an application can be described as a dictionary $\mathbf{D} \in \mathbb{R}^{N \times C}$, where $C$ is
the set of performance events collected across several configurations. 
Mathematically, this assumption is
expressed as follows: given a target (e.g., runtime, device utilization), $\mathbf{t} \in
\mathbb{R}^{N}$, where $N$ is the number of runs, and the sparse representation, $\mathbf{a} \in
\mathbb{R}^{C \times 1}$ can be obtained as
\begin{equation}
\min_{\mathbf{a}} \|\mathbf{t} - \mathbf{D} \mathbf{a}\|_2^2 \text{ s.t. } \|\mathbf{a}\|_0 \leq \kappa
\label{eqn:spcode}
\end{equation}
\noindent
where $\|.\|_0$ denotes the $\ell_0$ norm that counts the total number of
non-zero entries in a vector, $\|.\|_2$ is the $\ell_2$ norm, and $\kappa$ is
the desired sparsity, which is a user-defined variable. The higher the value of $\kappa$, the sparser solutions become. For all experiments presented in this paper, we use $\kappa$=0.5, which means 50\% sparsity.
This optimization problem solves for the
the sparsest set of performance metrics that can predict $\mathbf{t}$. 
Exact determination of the sparsest representation using Equation~(\ref{eqn:spcode})
is an NP-hard problem~\cite{davis1999combined}, and hence it is common to adopt greedy pursuit techniques such as \textit{Orthogonal Matching Pursuit} (OMP)~\cite{bruckstein2008sparse}. 
OMP algorithm selects the hardware event that is most strongly correlated to the target, removes its contribution, and iterates. This approach makes a sequence of locally optimal choices to approximate the globally optimal solution. 

Since a selective approach such as OMP misses opportunities to identify all critical factors in explaining a target performance metric, Islam et al. extended the OMP methodology to create an ensemble.
Instead of choosing the most correlated dictionary element in
each step, \textit{Ensemble OMP} selects the top $\tau$ most correlated metrics (i.e., similar coefficients), creates a discrete probability distribution, and randomly selects one using this distribution. For all experiments presented in this paper, we use $\tau$=5, meaning if several performance events are strongly correlated to each other in the way they drive a target, the algorithm would consider the top 5. It then predicts the target using the selected event, computes the residual, and repeats this process until the desired sparsity is met or the model achieves sufficient fidelity. This randomized algorithm repeats $50,000$ times independently (a user-defined parameter in the \dash framework). The final representation is obtained by averaging the solutions in the ensemble. 
\subsection{Computing Resource Importance}
The greedy approach selects each performance event, predicts the target metric, computes the error in prediction using Equation~\ref{eqn:eu_eqn5}. The error is then converted into a converse measure called \textit{the belief} using Equation~\ref{eqn:eu_eqn6} such that if the prediction error is low, an event is believed to be important. The Ensemble OMP algorithm aggregates the beliefs of all of the events in a resource group to compute the Resource Significance Measure (RSM) of an entire resource using Equation~\ref{eqn:rsm}. A resource with high RSM significantly influences a target performance metric. 
\begin{gather}
\boldsymbol{e}_i = \|\boldsymbol{t} - \boldsymbol{d}_ia_i\|_2^2
\label{eqn:eu_eqn5}\\
\boldsymbol{\alpha}_i = \exp\left(-\gamma \times \boldsymbol{e}_i \right)
\label{eqn:eu_eqn6}\\
\mathit{RSM}_r^w = 1 -\prod_{i \in L_r} (1 - \boldsymbol{\alpha}_i)
\label{eqn:rsm}\\
\mathit{RSM}_r = \frac{1}{W} \sum_{w=1}^W \mathit{RSM}_r^w
\label{eqn:rsmw}
\end{gather}
In Equation~\ref{eqn:rsm}, $(1 - \boldsymbol{\alpha}_i)$ describes the inability of the $i^{th}$ event of resource $r$ to perfectly predict the target metric. Hence, the term $\prod_{i \in L_r} (1 - \boldsymbol{\alpha}_i)$ describes the deficiency in predicting the target metric $t$ considering all events in the resource group $r$; this means $1-\prod_{i \in L_r} (1 - \boldsymbol{\alpha}_i)$ describes the combined ability of all of the events in successfully predicting the target metric for a specific workload $w$. Equation~\ref{eqn:rsmw} averages the RSM of a resource $r$ for all input parameters.

\section{Our Approach}
\label{sec:overview}
This paper aims to \textit{characterize common optimizations regarding interactions with hardware components and how they impact an individual or composite target metric.} Specifically, we design composite target metrics to capture the trade-off space between execution time and device utilization since they are often inversely correlated. Then, we characterize the optimizations based on how they influence the trade-off space and then leverage the characteristics to suggest optimizations to improve the trade-off space. The overall approach of this research involves three steps--(1) measurement, (2) analysis, and (3) visualization. The following sections present more details about the composite metrics and methodologies we developed in this work. While we focus on \nvidia GPUs in this work, the metrics and methods developed in this paper are generic. 

\subsection{Measurement}
\nvidia provides two distinct ways of collecting device and application monitoring information. 
\nvidia Management Library (NVML)~\cite{nvml} provides a transparent, non-intrusive mechanism to collect high-level device-oriented metrics (e.g. GPU utilization). \nvidia CUDA Profiling Tools Interface (CUPTI) provides a fine-grained, in-source performance monitoring interface~\cite{cupti-ref}. We use the NVML and CUPTI profiling capabilities in two ways. 
First, we use our thread-based profiling library \libpmon 
to collect NVML metrics separately for each CUDA kernel invocation. \libpmon provides a lightweight in-source annotation interface to demarcate the start and end of each CUDA kernel. \libpmon initiates a sampling thread on the host CPU, which synchronously starts and stops profiling at \texttt{start\_sample} and \texttt{end\_sample} calls, respectively.   
Second, for the automatic and low-overhead collection of CUPTI event counters, we employ the~\tool~\cite{libnvcd} tool. 
We prefer~\tool over other profiling tools~\cite{shende2006tau,adhianto2010hpctoolkit} due to its programmability, better control over measurements, and granularity. Specifically,~\tool provides source-level annotations to enable selective, per-kernel CUPTI profiling and reporting\footnote {In CUDA 11.0.2, \nvidia allows the developers to collect performance data only in kernel-level granularity }. \tool also provides an interface to automatically identify hardware performance events that can be monitored together without incurring additional overhead, which alleviates the need to make any ad hoc groups. The~\tool tool intercepts a selected list of CUDA kernels. 

\subsection{Analysis}
\label{sec:perf-metric}
This paper builds on an open-source performance analytics framework \dash~\cite{islam2019toward} and extends its capabilities (described in Section~\ref{sec:background}) to achieve the stated objective. 
Identifying important hardware performance events to explain device utilization and execution time can be formulated as a feature selection task in machine learning. Based on sparse coding theory, an application's performance metric can be decomposed into a small set of elementary patterns. These patterns are drawn from a dictionary of performance events collected under hundreds of configurations.

Hardware performance events capture application-device interactions. Several such interactions (e.g., misses, hits, reads, stalls) can occur for each high-level hardware resource (e.g., L1 cache). Using the methodology described in Section~\ref{sec:background}, 
we characterize GPU code optimizations based on their impact on the resource usage and, in turn, on application performance and SM utilization. To study the joint-optimization problem, we design a composite metric that is a function of both. The more closely this function captures the relationship between execution time and device utilization, the more effective it will be in distinguishing application characteristics. The other naive approach of characterizing optimizations concerning execution time and SM utilization separately does not lead to a quantitative measure to understand the joint trade-off space. In comparison, a joint target metric provides a unified method of studying the trade-off space without making manual comparisons.

\noindent\textbf{Proposed Target}
We set up feature selection as a minimization problem where features that reduce the target vector's reconstruction error are selected. The rationale for posing the question as a minimization one is that the identified application-device interactions are the ones that need to be reduced to decrease utilization loss ($1-ul$ in Equation~\ref{eqn:eu_eqn4}) or improve application performance ($ts$ in Equation~\ref{eqn:eu_eqn4}) or optimize both. 

\begin{gather}
\vspace{-0.5in}
target \rightarrow 
\begin{cases}
  ts & \text{for application performance}\\    
  1-ul & \text{for SM utilization}\\    
  \score & \text{for both application performance} \\
  & \text{and SM utilization}\\    
\end{cases}
\label{eqn:eu_eqn4}
\vspace{-0.5in}
\end{gather}
Here, $ts$ is the normalized average kernel execution time ($0 \le ts \le 1$), $ul$ is the normalized average SM utilization ($ 0 \le ul \le 1$), and \score is defined as:
\begin{gather}
\label{eqn:eu_eqn3}
\score \rightarrow 1 - \frac{\alpha}{ts}, where \\    
\alpha \rightarrow 
\begin{cases}
  \alpha_1 & \text{for }0 \le ul < 0.5 \text{ meaning low utilization}\\    
  \alpha_2 & \text{for }0.5 \le ul < 0.8 \text{ meaning moderate utilization}\\    
  \alpha_3 & \text{for }0.8 \le ul \le 1.0 \text{ meaning high utilization}\\  
\end{cases}
\vspace{-0.4in}
\end{gather}
Our empirical studies in Section~\ref{sec:matmul-util} show that the salient features identified as the causes for low SM utilization may not be the same as that for explaining the execution time of a kernel. Hence, conducting these analyses with $1-ul$ and $ts$ as targets separately makes a quantitative comparison between their characteristics difficult.

To address this challenge, we propose a multi-objective metric, namely--\score (Equation~\ref{eqn:eu_eqn3}), that is a function of both $ul$ and $ts$.
We define $\alpha$ in Equation~\ref{eqn:eu_eqn3} as a function that categorizes utilization into three buckets (Equation~\ref{eqn:eu_eqn3}) (low, medium, and high). The rationale for transforming utilization into buckets is to ensure that the trade-off for configurations with \textit{similar} SM utilization (low, moderate, or high) leans towards minimizing execution time. Different iterations with different $\alpha_1$ and $\alpha_2$ values reveal that as long as $\alpha_1$, $\alpha_2$, and $\alpha_3$'s value is within the corresponding $ul$ range, the exact value of those do not affect the analysis that much. For the experiments in this paper, 0.1, 0.5, and 0.8 are set as $\alpha_1$, $\alpha_2$, and $\alpha_3$'s values, respectively.

\subsection{Proposed Comparative Analysis Approach}
\label{sec:comparative}
This paper also presents a comparative analysis approach for correlating the difference in resource usage behavior to that in $target$ between applications across configurations. For the comparative study, we propose to derive new features based on the same approach explained in Section~\ref{sec:background} by replacing $\delta D = D_{1} - D_{2}$ in the Equation~\ref{eqn:spcode} and computing RSM. Here, $D_{1}$ presents the performance dictionary for a kernel, application, or configuration, and $D_{2}$ presents the same from a second one. Then, the problem can be formulated as characterizing the impact of $\delta D = D_{1} - D_{2}$ on $\delta T = target_{1} - target_{2}$

We can map the problem of correlating performance change to that of hardware usage behaviors as a feature selection problem. The new features are the differences between the two kernels' performance events (e.g., baseline and optimized). Figures~\ref{fig:optimization-perf}b, d, e demonstrate the information that is calculated in this experiment. This methodology correlates both $\delta D = D_{1} - D_{2}$ and $\delta D' = D_{2} - D_{1}$ with $\delta T$ such that a strong correlation between $\delta D$ and $\delta T$ (tall bar in the negative direction) indicates that those resources are optimized less by code transformation with respect to the baseline. On the other hand, a strong correlation between $\delta D'$ and $\delta T$ (tall bar in the positive direction) indicates that the code transformation increases the use of that resource. This comparative study allows us to identify which optimizations improve compute resource utilization (e.g., a tall positive bar for PCIE) and which ones reduce expensive interactions with the memory subsystem (e.g., a tall negative bar for DRAM). Based on these observations, we then create several resource usage characteristics-based code transformation suggestions in Table~\ref{tab:opt-steps} (Briefly described in Section~\ref{sec:pennant}), which can help users with new applications to systematically apply a subset of these optimizations to improve execution time, SM utilization, or both on \nvidia Volta platform.


\section{Experimental Setup}
\label{sec:setup}
\subsection{System}
We implement and evaluate our approach on a heterogeneous supercomputer
with 792 compute nodes and four NVIDIA Volta GPUs per node, running at 3.1 GHz with a combined 253 TB memory. 
We use the NVML and CUPTI Application Programming Interfaces (APIs) of \nvidia CUDA 11.0.2 toolkit for GPU profiling and management. 
\nvidia Volta hosts a multi-tier memory system with a private L1 cache and a shared memory for each SM connected to a shared L2 cache and DRAM on-chip for the entire GPU. The experiments are run on one NVIDIA Volta GPU. 

\subsection{Applications}
We analyze two open-source ECP proxy applications and a benchmark designed to test the mixed-precision performance of GPUs. 

\noindent \textbf{\pennant} is a proxy application for hydrodynamics on general unstructured meshes in 2D (arbitrary polygons). It makes heavy use of indirect addressing and irregular memory access patterns.  It contains mesh data structures and a few physics algorithms adapted from the LANL rad-hydro code FLAG and gives a sample of FLAG's typical memory access patterns. We used a modified version of Leblancx and sedov input decks for our tests. For the Leblancx input, we varied the mesh sizes from 600x600 to 2100x2100 in steps of 500.

\noindent \textbf{\minife} is a proxy application for unstructured implicit finite element codes similar to HPCCG and pHPCCG with a complete vertical covering of the steps in this class of applications. It also supports computation on multicore nodes, including pthreads and Intel Threading Building Blocks (TBB) for homogeneous multicore and CUDA for GPUs. 

\noindent \textbf{\mixbench} is a benchmark developed to evaluate GPUs' performance bounds on mixed operational intensity kernels~\cite{mixbench}. The executed kernel is customized on a range of different operational intensity values. Modern GPUs can hide memory latency by switching execution to threads able to compute operations. Using this benchmark, one can assess the practical optimum balance in both types of operations for a GPU. 

\noindent \textbf{\matmul} is a synthetic benchmark to perform matrix-matrix multiplication, a kernel that is fundamental to scientific computing, with incremental optimization added to the baseline code~\cite{matmultref}. We incrementally apply the optimization steps to characterize. For input, we use square matrices of sizes 60x60 to 200x200 in steps of 5x5. 

Table~\ref{tab:parameters} presents a summary of the different input parameters varied to collect both high-level metrics and low-level counters for the four applications. 
\begin{table}[t]
\centering
\caption{Input configurations for the applications.}
\vspace{-0.1in}
\label{tab:parameters}
{\footnotesize
\begin{tabular}{ll} \toprule
  Name & 
  Input \\ \midrule
    \textbf{\pennant} & Leblancx grid sizes = 600x600, 600x1100, 600x1600, \\
  & 600x2100, 600x2600, 1100x600, 1100x1100, 1100x1600, \\ 
  & 1100x2100, 1100x2600, 1600x600, 1600x1100, 1600x1600 \\
  \hline
    \textbf{\minife} & s = 10, 60, 110, 160, 210, 260 \\ \hline

     \textbf{\mixbench} & Input = 1, 2, 4, 8, 16, 32 \\ \hline
  \textbf{\matmul} & 60x60, 65x65, 70x70, 75x75, 80x80, 85x85, 90x90, 95x95,\\& 100x100, 105x105, 110x110, 115x115, 120x120, 125x125,\\& 130x130, 135x135, 140x140, 145x145, 150x150, 155x155, \\& 160x160, 165x165, 170x170, 175x175, 180x180, 185x185, \\& 190x190, 195x195, 200x200. \\ 


    \bottomrule
\end{tabular}
}
\end{table}

\subsection{Defining Abstract Machine Model}
\label{sec:machine-model}
\begin{figure}[t]
  \centering
{\includegraphics[width=0.75\columnwidth]{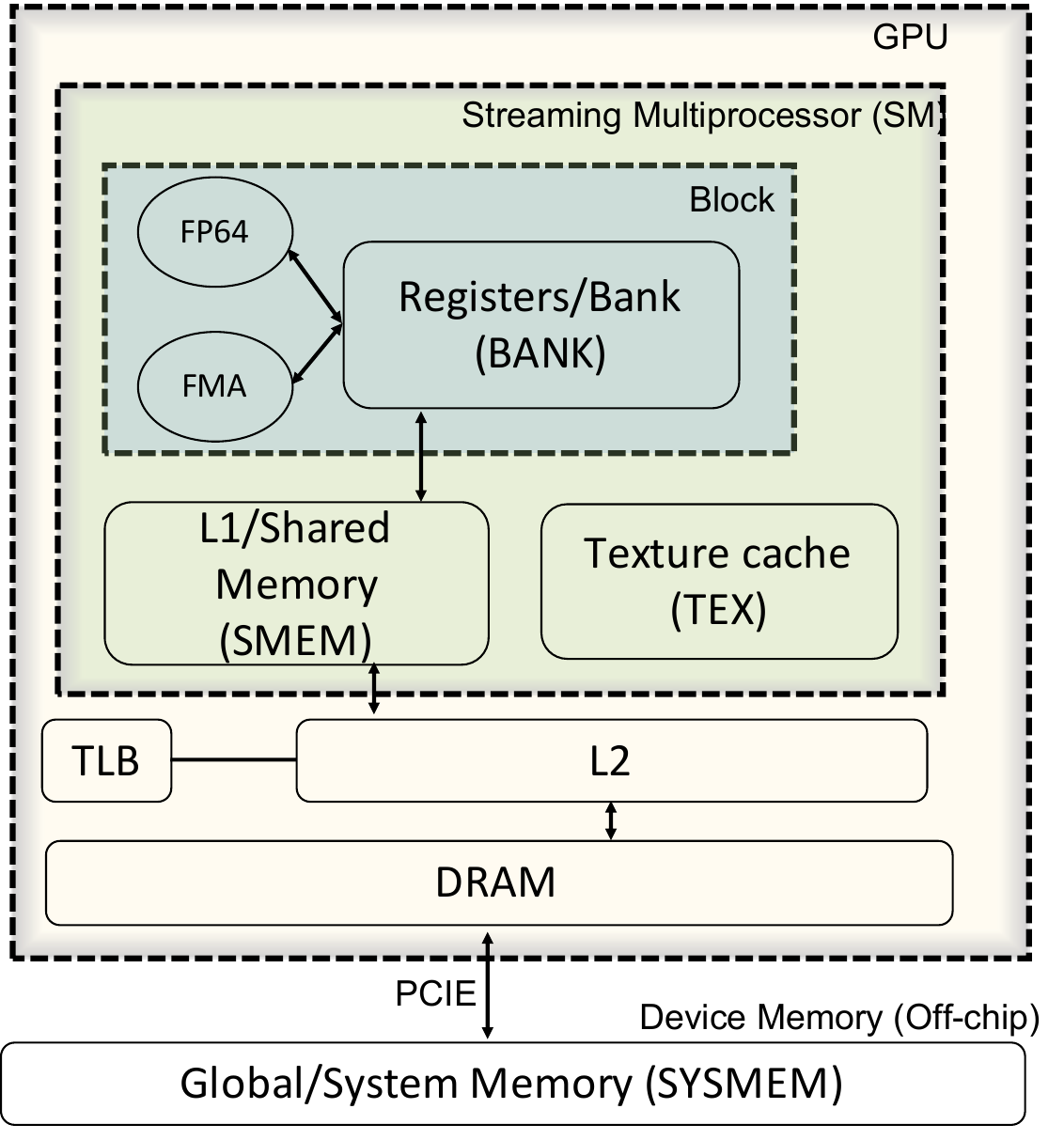}}
    \captionsetup{font=small}
    \caption{Abstract machine model of the \nvidia Volta architecture.}
    \label{fig:volta-arch}
\end{figure}
This section describes our approach of defining an abstract machine model for the \nvidia Volta GPU, presented in Figure~\ref{fig:volta-arch} (briefly described in~\ref{sec:machine-model}), that describes the hardware components and their scope (core-, chip-, node-level resource). An SM contains several Blocks that include computational units such as FP64 and FMA and registers called a Bank. Each SM consists of an L1, or shared memory cache shared across several blocks on the SM and provides a low latency memory space. Multiple SMs share the L2 and DRAM memory components on-chip, where DRAM is slower yet larger than L2. And finally, various GPUs on the same node share the Global/System memory space connected through the PCIE bus to the on-node resources. These components are the basis for our resource groups listed in Table~\ref{tab:resource-groups}.
The Volta architecture includes many more components than the ones presented in this machine model. However, those components either do not have any performance counters associated with them or define no way to access them. We develop a Python-based tool to automatically categorize the Volta performance events into the resource groups listed in Table~\ref{tab:resource-groups} using name-based prefix match and their meanings. For example, all counters that contain the sub-string BANK and are of type \ttt{\{read, write, queries, load, store\}} get categorized into the BANK group. However, the performance events describing cache or memory misses (e.g., \ttt{fp\_p0\_write\_misses}) need special consideration since these events incur latency that is at least equal to the cost of accessing the next level (slower but bigger) in the memory hierarchy. Hence, cache miss events are attributed to the next level in the memory hierarchy.
Additionally, four performance events do not contain the associated resources' prefix, which we manually resolved based on the interactions they capture.
The high-level groups (hardware resources) and the low-level factors (hardware performance events or application-device interactions) are configurable and change with different architectures. We will make the resource mapping available through the \dash framework on Github.

Table~\ref{tab:resource-groups} presents the high-level resources and the hardware performance events (application-device interactions) categorized into each group. We describe multiple events using regular expressions such that $[read, write]$ indicates read or write operations, and $*$ indicates any match. For this analysis, we exclude the \ttt{hit} events since such events do not describe a performance problem. Additionally, it is common knowledge that computational resource utilization (e.g., FMA, FP64) improves performance while that related to the memory hierarchy is detrimental. Hence, the computational resource usage should be \textit{maximized}, while others should be \textit{minimized} to gain better performance. For brevity, we use the term \textit{optimize} to refer to both, where the meaning will depend on the resource in the discussion being a computational or memory-related one. In the evaluation section (Sections~\ref{sec:optimizations} and~\ref{sec:results}) we shorten the event names where $r \rightarrow read$, $w \rightarrow write$, $q \rightarrow queries$, $p* \rightarrow subp*$, and $a \rightarrow active$. Here, $subp0$ and $subp1$ indicate two sub-partitions of the L2 resource; $fb\_subp*\_[read|write]\_[queries|sectors]$ indicates the number of DRAM read or write requests to the two sub-partitions of DRAM. Note, we categorize $fb\_subp*\_[read|write]\_misses$ as a SYSMEM event since a DRAM read or write that misses DRAM will access SYSMEM and will incur the latency of SYSMEM access.
\begin{table}[h]
\vspace{-0.1in}
\centering
\caption{Resource groups on \nvidia Volta}
\vspace{-0.15in}
{\footnotesize
\begin{tabular}{ll} \toprule
Resources	& H/W Events 	\\ \midrule
FP64 		& 
\begin{tabular}{l}
  inst\_executed\_fp64\_pipe\_[s0,s1,s2,s3] \\ \hline
\end{tabular}  \\
FMA &
\begin{tabular}{l}
  inst\_executed\_fma\_pipe\_[s0,s1,s2,s3] \\ 
  not\_predicated\_off\_thread\_inst\_executed \\\hline
\end{tabular}  \\
SMEM 		& 
\begin{tabular}{l}
  shared\_[ld,st]\_transactions \\ \hline
\end{tabular}  \\
TEX 		& 
\begin{tabular}{l}
  l2\_p[0,1]\_[read,write]\_tex\_sector\_queries \\ \hline
\end{tabular}  \\

BANK 		& 
\begin{tabular}{l}
  shared\_[ld,st]\_bank\_conflict \\ \hline
\end{tabular}  \\

L2 		& 
\begin{tabular}{l}
  l2\_p[0,1]\_total\_[read,write]\_sector\_queries \\\hline
\end{tabular}  \\

DRAM 		& 
\begin{tabular}{l}
  fb\_p[0,1]\_[read,write]\_[sectors,queries]\\  \hline
\end{tabular}  \\
SYSMEM 		& 
\begin{tabular}{l}
  fb\_p[0,1]\_[read,write]\_misses\\
  global\_[atom,cas,load,store] \\ 
  l2\_p[0,1]\_[read,write]\_sysmem\_* \\\hline
\end{tabular}  \\ 
PCIE 		& 
\begin{tabular}{l}
    pcie\_[rx,tx]\_active\_pulse\\ 
\end{tabular}  \\
\bottomrule
\end{tabular}
}
\label{tab:resource-groups}
\vspace{-0.1in}
\end{table}

\subsection{Extensions to~\dash}
\label{sec:extensions}
This section describes how we implemented our metrics and methodology in an open-source performance analytics framework--\dash.
\begin{figure}[t]
  \centering{\includegraphics[width=\columnwidth]{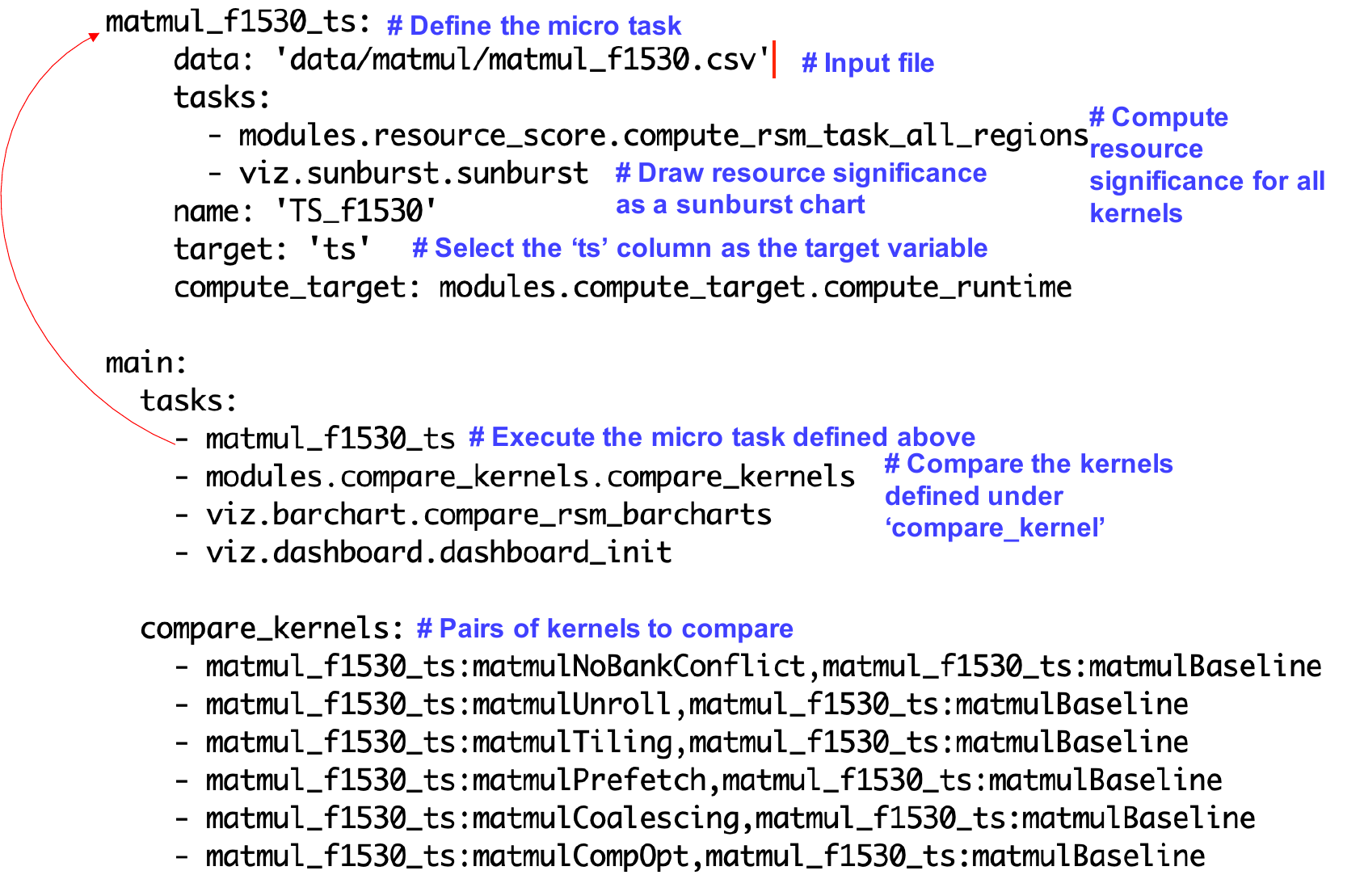}}
  \vspace{-0.25in}
  \caption{Setting up the~\dash framework.}
  \label{fig:dash-yml}
  \vspace{-0.15in}
\end{figure}
Figure~\ref{fig:dash-yml} shows an example of how to setup analysis and visualization tasks in the~\dash framework. As shown in Figure~\ref{fig:dash-yml}, the input data file indicated by \ttt{data} lists all kernels of an application (e.g., \matmul has seven kernels) for various configurations (e.g., input parameter). The \ttt{micro\_task} (\ttt{matmul\_f1530\_ts}) defines the analysis for a selected target variable using the \ttt{target} key (e.g., \ttt{ts}), which corresponds to one of the \ttt{target} fields. We implement a custom pre-processing function \ttt{modules.compute\_target.compute\_runtime} to normalize the selected \ttt{target} variable across all kernels and configurations. 
We implement the comparative analysis technique as a separate analysis in the \ttt{modules.compare\_kernels} function and the corresponding visualization in the \ttt{viz.barchart.compare\_rsm\_barcharts} function. The \ttt{compare\_kernels} section defines the pairs of kernels that are compared in this form: \ttt{$<$task1:kernel\_name1,task2:kernel\_name2$>$}, where \ttt{task1} and \ttt{task2} refer to micro tasks defined earlier. This comparative analysis is referred to as a `multi-source analysis' in the \dash framework. The \ttt{compare\_kernels} correlates the changes in resource usage behaviors between \ttt{task1} and \ttt{task2} to the change in the target performance metric.

\subsection{Visualization}
\begin{figure*}[t]
  \centering
{\includegraphics[width=\textwidth]{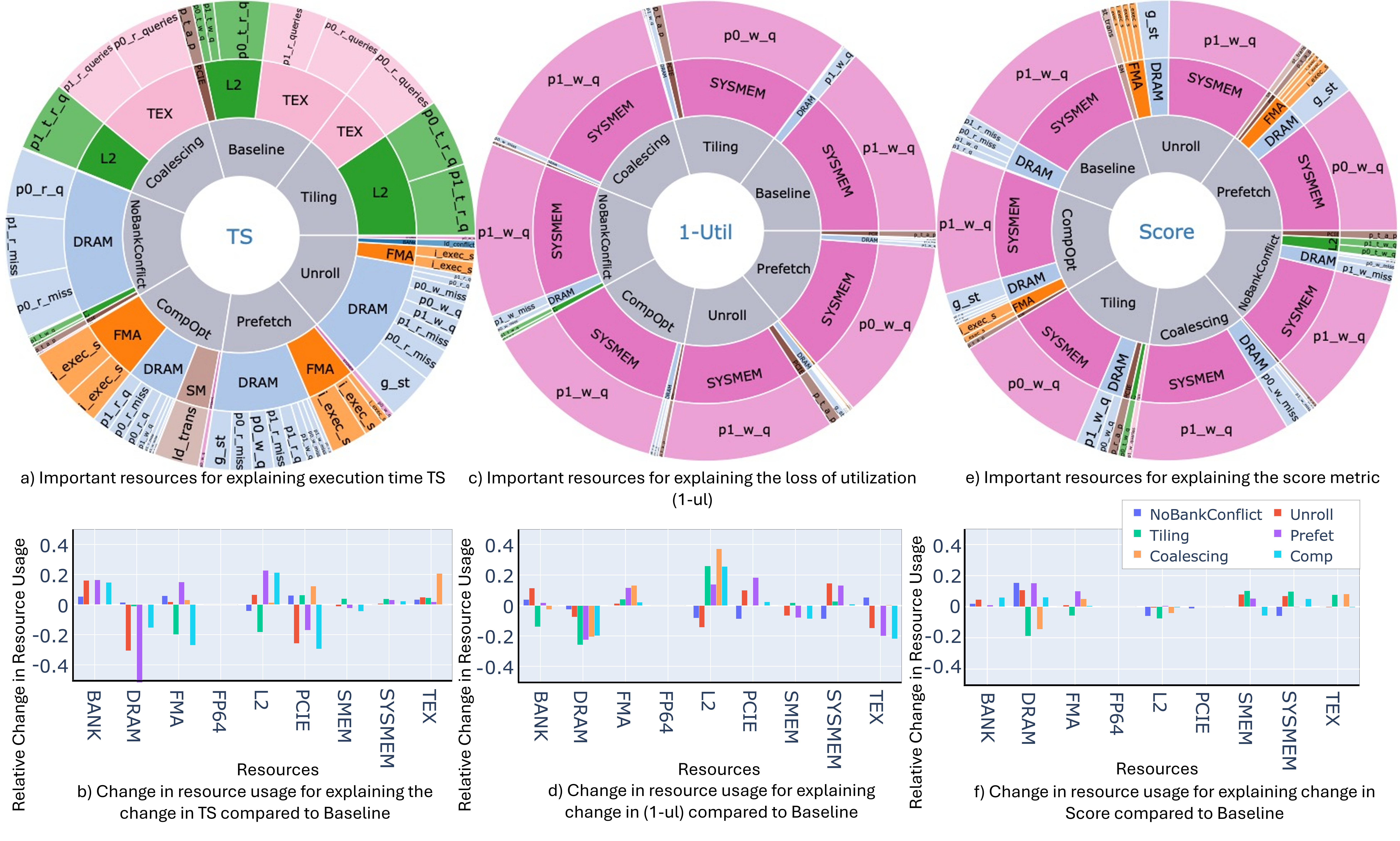}}
  \caption{Hardware resource usage behaviors that explain different $target$ metrics for various \matmul kernels.}
  \label{fig:optimization-perf}
\end{figure*}

We present hardware resource usage behaviors of applications in a hierarchical manner. 
Moving outwards on the sunburst chart (e.g., Figure~\ref{fig:optimization-perf}c) (results from Figure ~\ref{fig:optimization-perf}c described in \ref{sec:optimizations}), the first inner circle represents the entire application; the next circle represents each of the GPU kernels, then critical resources to explaining the $target$ of the parenting region across configurations, and finally, the important hardware counters (application-device interactions) that contribute to its respective resource. The sunburst chart shows each kernel's resource usage characteristics across several inputs and configurable parameters, such as frequency and power cap. The usage of color to represent individual resource groups and events is to differentiate between each element clearly; apart from that, it does not have any other implications. In this paper, we characterize kernel behaviors that persist across the input gamut. We also present the comparative resource usage behaviors of optimizations compared to the baseline in an easy-to-understand bar graph (e.g., Figure~\ref{fig:optimization-perf}b, d, f).

\section{Hardware Resource Usage of Optimizations}
\label{sec:optimizations}


\addtolength{\tabcolsep}{-0.4em}
\begin{table}[t]
\centering
\caption{Summary of optimizations.}
\vspace{-0.1in}
\label{tab:optimizations2}
{\footnotesize
\begin{tabular}{lll} \toprule
  Optimizations & 
  Descriptions & 
  \begin{tabular}{l}
    Resource Usage \\ Behaviors
  \end{tabular} \\  \midrule
    Tiling & We decompose the output & TS=\{L2, TEX\}\\ & matrix (C) into tiles and & 1-UL=\{SYSMEM\}\\&compute each output tile & Score=\{SYSMEM\\ &using corresponding input &,DRAM\}\\ &tiles. The kernel reads tiles\\ & of input matrices from\\ & DRAM to shared or local\\ & memory once and computes\\ & the output tile by iterating\\& over them. \\  \hline

    ComputeOpt &Inner product between A and B & TS=\{FMA, SM,\\ & tiles requires two operands, & DRAM\} \\ & however current streaming & 1-UL=\{SYSMEM\}\\ & multiprocessors only allow one & Score=\{SYSMEM,\\ & from the shared memory. Hence, & DRAM,FMA\}\\ &  we apply the outer product to \\ & A and B tiles, where A resides in\\ &  shared memory, but matrices B \\ & and C in registers. This \\ & transformation reduces the \\ & number of instructions \\ \hline

     Unroll & Add \#pragma unroll to explicitly  & TS=\{FMA,\\ & unroll the outer loop of the & DRAM\}\\ & Matrix-matrix multiplication & 1-UL=\{SYSMEM\}\\ & operations. By default, compilers & Score=\{SYSMEM, \\ & unroll the inner loops. & DRAM,FMA\}\\ \hline

    Coalescing & Column-major access to the & TS=\{L2, TEX\}\\ & input matrix B during tiling is & 1-UL=\{SYSMEM\}\\ & inefficient. B can be transposed & Score=\{SYSMEM\}\\ & by CPU before offloading it to \\ & GPU memory.\\ \hline

    NoBankConflict & Coalescing memory results in & TS=\{DRAM\}\\ & bank conflict, where different & 1-UL=\{SYSMEM,\\ & threads stall due to overlapping & DRAM\}\\ & memory access. By changing the  & Score=\{SYSMEM,\\ & order of access to B loaded in the & DRAM, L2\}\\ & shared memory, thread blocks \\ & can access the shared memory \\ & banks simultaneously given \\ & no dependency\\ \hline
    \bottomrule
\end{tabular}
}
\vspace{-0.2in}
\end{table}

\label{sec:matmul-util}

\begin{figure}[t]
  \centering{\includegraphics[width=\columnwidth]{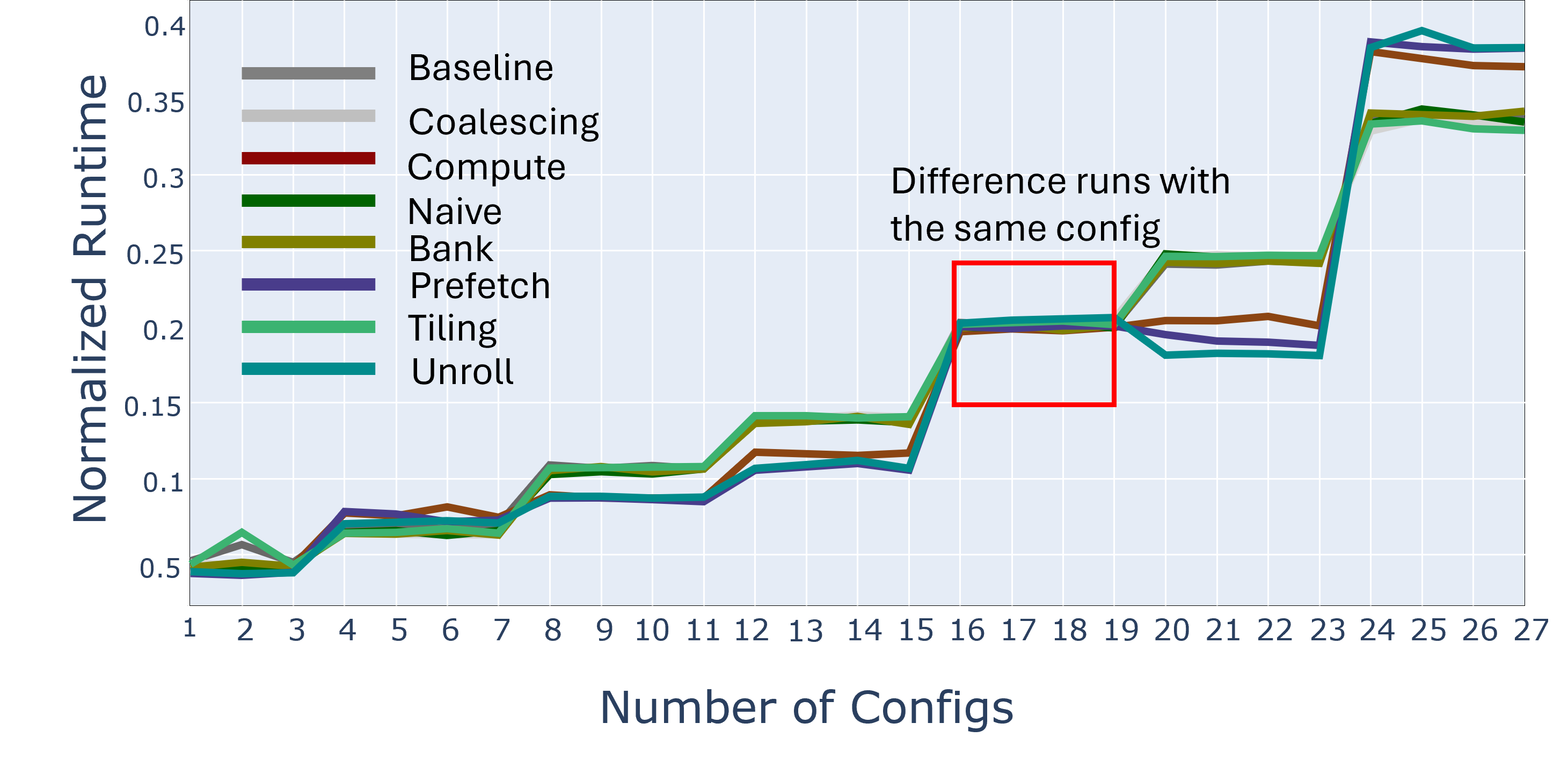}}
     \vspace{-0.25in}
    \caption{Execution times of all seven \matmul versions.}
    \captionsetup{font=tiny}
    \label{fig:matmul_ts}
  \vspace{-0.2in}
\end{figure}
This section thoroughly studies the resource usage characteristics of several common optimization strategies and their impact on a multi-objective space. Specifically, we explore six different optimizations for the \matmul application to quantify how their resource usage behaviors change compared to the baseline. 
We choose the \matmul application since it is one of the most widely used dwarfs in scientific and data analytics applications.
We analyze the following kernels --(1) \baseline (without optimization) (2) \tiling (3) global memory coalescing (\coalescing), (4) reducing shared memory bank conflicts (\bank), (5) computation optimization (\compute), (6) loop unrolling (\unroll), and (7) prefetching (\prefetch). This list is non-exhaustive because this paper aims to demonstrate the process of systematically identifying optimizations to maximize a multi-objective target metric, not to characterize all possible optimizations. The steps of such a process would be to (1) characterize a new application based on a target metric, and (2), depending on the important resource, identify a transformation that optimizes that target metric. 
Table 3 describes selected transformations' essence from a high-level and the significant resource usage behaviors they demonstrate. From Table 3, we can observe that the \score variable identifies the resources that are important for both $ts$ and $ul$, which signifies that the proposed composite metric is sufficient to identify resource usage behavior that should be targeted for optimization to improving both $ts$ and $ul$. For more details on the code changes to incorporate these optimizations to other scientific applications, we refer to the literature~\cite{matmultref}. 
To our knowledge, this paper is the first to present a detailed comparison of optimization characteristics concerning achieving multiple objectives (improving execution time and streaming multiprocessor usage). 
The following sections present our observations using \ttt{ts} and \ttt{ul} as targets separately for the \matmul application and evaluating the effectiveness of using the proposed \score metric capturing the similar behaviors.

\subsection{Impact on Application Performance (TS)}
\label{sec:matmul-ts}
This section presents the resource usage behaviors of the transformations performed on the \matmul application described in~\cite{matmultref} and how those interactions affect execution time alone.
The $target$ variable for this experiment is execution time (\ttt{ts} in Equation~\ref{eqn:eu_eqn4}). The rationale is that optimization efforts should reduce the identified hardware event counts that can predict on the increase in execution time as the input size grows. 

Figure~\ref{fig:matmul_ts} presents the normalized execution times of the seven kernels to demonstrate their relative ranking concerning performance (execution time). We can observe that \unroll, \prefetch, and \compute perform better than the other optimizations at smaller input configurations of (configs $1-24$), but the other optimizations perform better larger input sizes ($200\times200$).

Figure~\ref{fig:optimization-perf} presents the resource usage behaviors that explain the application performance across multiple runs and input parameters. The importance (or RSM) values of the hardware performance events are normalized for each kernel separately to sum up to one. Hence, the RSM of the same resource cannot be compared directly across kernels using Figure~\ref{fig:optimization-perf}a. 

To facilitate a comparative study across optimizations, we present a multi-source analysis in Figure~\ref{fig:optimization-perf}b showing the resource usage behaviors of each optimized kernel compared to the ~\baseline. Figure~\ref{fig:optimization-perf}b explains how the resource usage behaviors change due to optimizations compared to the baseline kernel. 
The X-axis shows the different hardware resource groups, and the Y-axis presents the relative change in resource usage. $+0.5$ in the comparative bar chart (e.g., Figure~\ref{fig:optimization-perf}b) means $50\%$ more usage by an optimization compared to the baseline, and $-0.5$ means $50\%$ reduction in interaction with that resource. When a code transformation reduces a high-latency resource usage such as DRAM, L2, or SYSMEM or increases low-latency resource usage, that information becomes evident in this simple yet powerful visualization. The visualization explains \textit{why} the code transformation improves performance. This information provides the foundation for Table~\ref{tab:opt-steps} (Briefly described in Section~\ref{sec:pennant}), which lists code transformation suggestions to be applied based on GPU resource usage. 
Additionally, we can make several observations from Figure~\ref{fig:optimization-perf}.

\noindent\textbf{Observation 1: } Figure~\ref{fig:optimization-perf}a shows that the number of read and write queries and misses to the L2 memory (for both P0 and P1 sub-partitions) and access to the texture cache explain the increase in execution time for the~\baseline kernel as input size increases. 

\noindent\textbf{Observation 2: } The~\bank optimization ensures that no two threads access the same bank ($128$ byte registers) during the same read or write instruction by increasing the bank width for shared memory. This transformation improves shared memory bank alignment, reduces bank conflicts, and enables threads to load (store) data in parallel. The rest of the performance behavior of the~\bank kernel can be explained by the DRAM usage due to compulsory or capacity misses. Figure~\ref{fig:optimization-perf}b also shows that compared to the~\baseline,~\bank slightly reduces resource usage such as L2 but improves FMA and PCIE usage.  

\noindent\textbf{Observation 3: } The~\tiling optimization increases the computation-to-memory ratio by breaking the computation of the output matrix into several tiles where one thread block computes each tile. The operations include reading input matrices to the shared memory from DRAM, calculating the inner product for a tile, accumulating results, and writing the result from shared memory to DRAM. From Figure~\ref{fig:optimization-perf}a, we can observe that the \tiling kernel utilizes the Texture cache (where tiles reside, $l2\_p0\_read\_tex\_sector\_queries$) and the L2 cache (where A and B are initially read in).

\noindent\textbf{Observation 4: }The~\coalescing kernel fuses several memory locations reads into one by leveraging the knowledge of data access pattern of the~\matmul application (B is read along the column) and the observation that C/C++ manages memory in row-major order. By fetching needed data into the local cache, the~\coalescing kernel can utilize the Texture cache better ($l2\_p*\_read\_tex\_sector\_queries$). From Figure~\ref{fig:optimization-perf}b, we can see that the performance of the~\coalescing kernel improves over the~\baseline by improving Texture cache usage.

\noindent\textbf{Observation 5: }The~\compute kernel further optimizes the~\tiling kernel by reading the matrix A into the shared memory from DRAM. Since all threads share this matrix, it reduces the overall DRAM usage and improves SM (shared memory) usage. Since the \compute kernel computes A and B matrices' outer products with only one instruction (Table 3), it increases the compute-to-memory ratio (FMA).

\noindent\textbf{Observation 6: }The~\prefetch and~\unroll optimizations both apply prefetching of blocks of data, and the~\unroll optimization additionally unrolls the outer loop (Table 3). Hence, both of these kernels mainly use the shared memory cache for loading data (increase in $ld\_transactions$ in SMEM) to decrease the read and write L2 misses ($p0\_r\_miss$) that get fulfilled by the DRAM. The compute-to-memory ratio improvement is evident by the increase of FMA usage measured by the number of double-precision floating-point operations executed. 

\noindent\textbf{Observation 7: }Since~\prefetch,~\unroll, and~\compute kernel optimizations aim towards improving the compute-to-memory ratio, all three of the kernels exert similar resource usage behaviors. Figure~\ref{fig:matmul_ts} shows that even though their performances are significantly better (lower TS) compared to others at smaller input sizes (where the workload fits in the shared memory cache), improving the compute-to-memory ratio alone is not sufficient at large input since the workload may not fit into the shared memory cache. Hence, additional optimizations such as~\bank and~\coalescing may be required.

\noindent\textbf{Observation 8: } Figure~\ref{fig:optimization-perf}b presents information about the comparative study of all kernels to the ~\baseline. We can observe that the~\unroll,~\prefetch, and~\compute optimizations significantly reduce the L2 usage compared to the~\baseline by increasing compute-to-memory ratio, which manifests as high FMA usage.
\begin{figure*}[t]
  \centering
{\includegraphics[width=\textwidth]{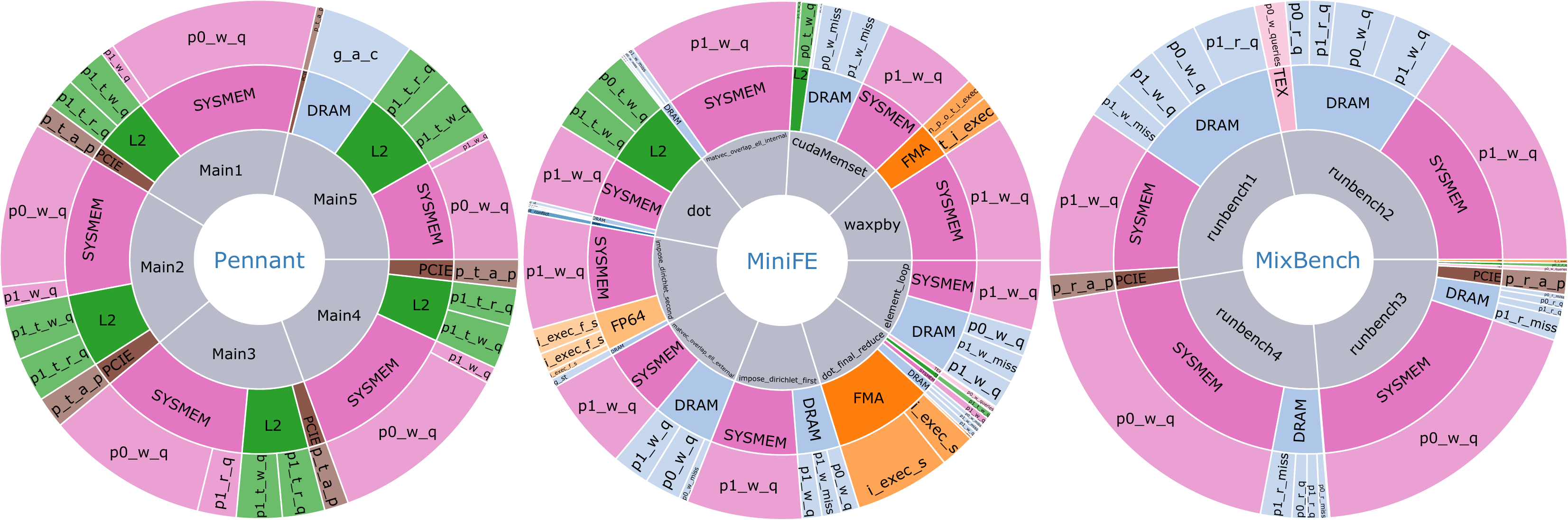}}
  \caption{Hardware resource usage behaviors that explain the $target \rightarrow \score$ of two ECP applications and a benchmark. Based on their resource usage, we propose and implement several optimizations that improve the performance of the \ttt{Main3} kernel in \pennant up to 29.6\% and SM utilization up to 5.4\%. }
  \label{fig:miniapps}
  \vspace{-0.1in}
\end{figure*}

\subsection{Impact on SM utilization}
Figure~\ref{fig:optimization-perf}c presents the hardware resource usage behaviors that impact SM utilization. This experiment aims to identify the hardware performance events that explain the loss of SM utilization. The rationale for selecting $target \rightarrow 1 - ul$ is that by identifying the application-device interactions that demonstrate why the device is not fully utilized, application scientists and performance analysts can design optimizations to reduce those expensive interactions to improve average SM utilization. 

Figure~\ref{fig:optimization-perf}c shows that the number of write queries ($write\_sysmem\_sector\_queries$ or $p*\_w\_q$) to the remote system memory explains the loss of SM utilization for all kernels. This behavior can be explained by the fact that while the device stalls for data transfer from (to) remote memory (SYSMEM), the compute unit sits idle.
Similarly, from Figure~\ref{fig:optimization-perf}d, we can observe that while almost all transformations but \tiling reduces DRAM traffic compared to the \baseline, which indicates improved SM utilization (not shown in the chart).

\subsection{Impact on both TS and SM utilization}
This study aims to identify the application-device interactions that can be reduced to improve both the execution time of applications (decrease) and SM utilization (increase). In this experiment, we use $target \rightarrow \score$ (Equation~\ref{eqn:eu_eqn3}) and identify the resource usage behaviors that can explain an increase in the \score (either low SM utilization or high application time). 

From Figure~\ref{fig:optimization-perf}e, we can observe that the \score metric identifies SYSMEM, measured by the number of reads/write to the remote memory, to have the most significant influence on performance and SM utilization. Further investigation of the normalized execution time and SM utilization graphs (not presented in the paper due to page limit) shows that the change in execution time between different kernels is relatively small and stable. However, SM utilization varies drastically, primarily due to the device stall due to SYSMEM access. Hence, characteristics of the \score chart lean toward that of the SM utilization. 

Figure~\ref{fig:optimization-perf}f shows that reducing the number of write queries to the system memory and DRAM while improving FMA and Texture cache usage will significantly allow the device to avoid stalling and improve the compute-to-memory ratio. From Table 3, we can observe that the $score$ metric captures the important resources identified based on Figure~\ref{fig:optimization-perf}a ($target \rightarrow ts$) for those optimizations that impact execution time more (\compute, \unroll, and \prefetch) by improving compute to memory ratio. Further investigation shows that these optimizations do not impact SM utilization much. Hence, the $score$ metric ``represents" the best interest of improving application performance. In contrast, the $score$ metric captures the important resource for $target \rightarrow (1-ul)$ (Figure~\ref{fig:optimization-perf}c) for those optimizations that improve SM utilization by reducing idle cycles due to stall for reading (writing) system memory. 
These optimizations do not negatively impact application performance as well. 
In summary, these observations show that instead of separate analyses for $ts$ and $1-ul$ as targets, it is sufficient to use the proposed $score$ metric to identify important resource usage behaviors.


\section{Optimization Opportunities for Miniapps}
\label{sec:results}
\begin{table*}[!ht]
\setcounter{table}{3}
\centering
\caption{Code tuning opportunities and optimization suggestions inferred from GPU resource usage.}
\label{tab:opt-steps}
\resizebox{\textwidth}{!}{%
\begin{tabular}{|c|c|c|c|c|}
\hline
Type &
  \begin{tabular}[c]{@{}c@{}}Impacted Resource \\ Group(s)\end{tabular} &
  \begin{tabular}[c]{@{}c@{}}Tuning Opportunity\end{tabular} &
  Suggested Code Transformation &
  \begin{tabular}[c]{@{}c@{}} Expected \\ Improvement \end{tabular} \\ \hline
A &
  \begin{tabular}[c]{@{}c@{}}Global memory \\ (DRAM)\end{tabular} &
  \begin{tabular}[c]{@{}l@{}}Convert global memory \\ 
                             accesses to shared memory \\ 
                             accesses\end{tabular} &
  \begin{tabular}[c]{@{}l@{}}Replace references to vectors allocated in global memory \\ 
                             to shared vectors (using \texttt{\_\_shared\_\_} keyword). Copy-in and -out \\ 
                      to the global vectors only at the start and end of the kernel.\end{tabular} &
  Primary: Speedup \\ \hline
B &
  \begin{tabular}[c]{@{}c@{}}Bank conflicts\\ (BANK)\end{tabular} &
  \begin{tabular}[c]{@{}l@{}}Align vector sizes to \\ cache line width to \\ minimize bank conflicts\end{tabular} &
  \begin{tabular}[c]{@{}l@{}}Change the data type of the shared vectors to 8 bytes per \\ element (assuming 8-byte cache line size). Also call at initiation: \\ \texttt{cudaDeviceSetSharedMemConfig(}\texttt{cudaSharedMemBankSizeEightByte)} \end{tabular} &
  \begin{tabular}[c]{@{}l@{}}Primary: Speedup,\\ Secondary: Utilization\end{tabular} \\ \hline
C &
  \begin{tabular}[c]{@{}c@{}} Cache counters  \\ (SMEM, L2)  \end{tabular} &
  \begin{tabular}[c]{@{}l@{}}Fetch as much data \\ into L1 as possible at the \\ time of kernel execution\end{tabular} &
  \begin{tabular}[c]{@{}l@{}}Replace frequently used small functions into macros or \\
                             expand function code in-place to eliminate L1 cache flushes \\ at call boundaries.\end{tabular} &
  \begin{tabular}[c]{@{}l@{}}Primary: Speedup, \\ Secondary: Utilization\end{tabular} \\ \hline
D &
  \begin{tabular}[c]{@{}c@{}}Stalls \\ (FP64, FMA, SMEM) \end{tabular} &
  \begin{tabular}[c]{@{}l@{}}Minimize or remove frequent \\
                             conditional assignments with \\ 
                             sequential code\end{tabular} &
  \begin{tabular}[c]{@{}l@{}}Replace conditional vector updates with bit-wise operations \\ 
                             to achieve effective conditional assignment. These include
(but \\
                             are not limited to) the following 
                             macros described in Figure~\ref{fig:bitwise-opt}:\\ 
                             OPT\_IF\_LESS\_SET, OPT\_IF\_LESS\_DOUBLE2, OPT\_IF\_LESS\_SELECT 
                             \end{tabular} &
  \begin{tabular}[c]{@{}l@{}}Primary: Utilization,\\ Secondary: Speedup\end{tabular} \\ \hline
E &
  \begin{tabular}[c]{@{}c@{}} Computation \\ counters \\ (FP64, FMA) \end{tabular} &
  \begin{tabular}[c]{@{}l@{}} Pipelining of multiple element \\ computations in a loop per
                              iteration, \\ especially for simple loops with \\ low                              dependencies in loop block.
                             \end{tabular} &
  \begin{tabular}[c]{@{}l@{}}     Add `\texttt{\#pragma unroll n}' before a loop which does not \\ 
                             contain conditional statements (\texttt{n >= \# of iterations}) \\
                             or manually unroll the loop.                             
                             \end{tabular}  &
  \begin{tabular}[c]{@{}l@{}}Primary: Utilization, \\ Secondary: Speedup \end{tabular} \\ \hline

\end{tabular}%
}
\vspace{-0.1in}
\end{table*}

Based on the observations in Section~\ref{sec:optimizations}, we summarize the common characteristics of selected optimizations in Table~\ref{tab:opt-steps} concerning their dominant interactions (resource groups), and prescribe optimizations 
to reduce those interactions (tuning opportunity), examples of transformation that will allow a user to apply these optimizations to their code. 
In this section, we apply the prescriptions in Table~\ref{tab:opt-steps} to resource usage characteristics of three proxy applications---\pennant, \minife, and \mixbench. 
The objective of these experiments is to demonstrate how users can leverage our prescriptions 
to new applications and interpret the pattern-driven optimization suggestions. While the recommendations provided in this paper are specific to \nvidia Volta, this work motivates the need for such characterization studies on new platforms. Once done, the resource usage characteristics-driven optimization suggestions will provide users and auto-tuners a systematic method of applying code transformation.

\subsection{Identify Optimization Opportunities}
From Figure~\ref{fig:miniapps}, we make the following observations:

\noindent 
\textbf{\pennant:} The \pennant application's performance and SM utilization are bounded by its memory usage behaviors. 
Based on our observations in Section~\ref{sec:optimizations}, all five of the \pennant kernels will benefit primarily from converting as many global memory accesses as possible to shared memory accesses. If that happens in the majority, the kernels will further benefit from bank conflict mitigation (as bank conflicts will dominate the target metric).  
Additionally, to improve the compute-to-memory ratio, \pennant will likely benefit from the \prefetch and unroll optimization, which can be easily incorporated by application scientists. Other optimizations such as \tiling and the \compute may also improve the FMA usage but require algorithmic change and are less trivial.
    
\noindent
\textbf{\minife:}
The collective behavior of application performance and SM utilization for \ttt{cudaMemset}, \ttt{impose\_dirichlet\_first}, \ttt{dot}, \ttt{matvec\_overall\_*} kernels in the \minife application can be explained by their DRAM and SYSMEM usages since all of these kernels involve memory copy and transfer from the system memory and DRAM to the local cache. These kernels will benefit from using the shared memory and local texture caches through optimizations such as removing bank conflicts and using memory coalescing. The \ttt{waxpby}, \ttt{impose\_dirichlet\_second}, and \ttt{dot\_final\_reduce} kernels compute on the data being fetched by the other kernels and already uses the \unroll optimization and thus, utilizes more FMA (compute) resources. 

\noindent
\textbf{\mixbench:}
Finally, the \mixbench application shows that all four benchmarks depend heavily on sending write queries through the PCIE bus to the system memory, which results in low SM utilization and poor application performance. Optimizations such as memory coalescing and bank conflict reduction can improve both application performance and SM utilization. Other optimizations such as \prefetch and \unroll can improve the compute-to-memory ratio.

\begin{figure*}[t]
  \centering
{\includegraphics[width=\textwidth]{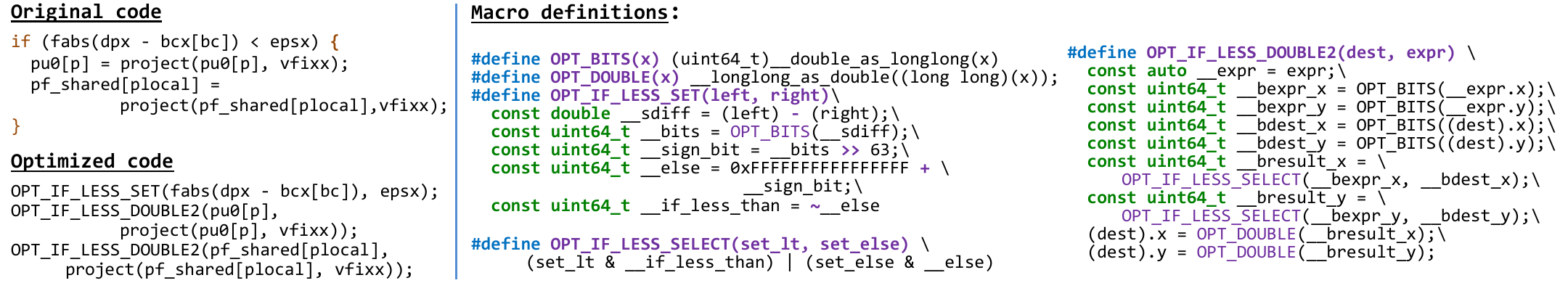}}
  \caption{Replacing conditionally executed operations on simple vectors with bit-wise operations.}
  \label{fig:bitwise-opt}
\end{figure*}

\subsection{Applying Optimization to \pennant}
\label{sec:pennant}
This section applies resource usage characteristics-driven optimizations to the \ttt{Main3} kernel in the \pennant application and measuring both execution time and SM utilization. This experiment's objective is to evaluate our approach's effectiveness, which maps resource utilization behaviors to potential optimizations. 
In this experiment, we incrementally apply optimizations described in Table~\ref{tab:opt-steps}. From Figure~\ref{fig:miniapps}, we can observe that the impact of the system memory(SYSMEM) on the \score metric is the largest. This implies that the \ttt{Main3} kernel of the \pennant application uses the system memory extensively, which results in higher latency. So, to reduce that usage, we replaced three global vectors \ttt{pmaswt}, \ttt{pf} and \ttt{pu0} by three similarly-sized shared memory vectors that we used during the execution of the kernel (Transformation 'A' in Table~\ref{tab:opt-steps}). At the end of the kernel, we updated the global memory buffers with their shared memory counterparts, thereby significantly reducing the system memory group's dominance on the kernel execution time. At this time, bank conflicts emerged as the dominant resource group (intermediate analysis not shown). To minimize bank conflicts, we then applied two different strategies: we replaced the 4-byte sized vectors (doubles) by 8-byte sized vectors (double2 for \ttt{pf} and \ttt{pu0}) for better bank access alignment. Additionally, we changed the distribution of shared memory with respect to the L1 cache size using \ttt{cudaDeviceSetSharedMemConfig} and \ttt{cudaDeviceSetCacheConfig} calls, respectively (Transformation 'B' in Table~\ref{tab:opt-steps}). These additional optimizations further reduced the dominance of bank conflicts on the kernel execution time. Finally, in order to improve the L1 cache behavior, we in-lined the following functions being called by \ttt{Main3}: \ttt{calcAccel()}, \ttt{advPosFull()}, \ttt{gatherToPoints()} and \ttt{applyFixedBC()} (Transformation 'C' in Table~\ref{tab:opt-steps}). 
We evaluated the optimization on two inputs of \pennant: Sedov and Lebancx. For Leblancx input, we observed an average 27.4\% improvement in execution time (maximum 29.6\%), an average 2\% improvement in SM utilization (maximum 5.3\%), and an average 12.2\% improvement in power draw (maximum 14.8\%). For Sedov input, we see an average execution time improvement of 19.7\% (maximum 23.4\%), an average reduction in power usage of 14\% (maximum 26.54\%), but minor improvement in SM utilization.



\section{Related Work}
\label{sec:related}
Several performance measurement toolkits provide monitoring and analysis capabilities for both CPUs and GPUs. HPCToolKit~\cite{adhianto2010hpctoolkit} and TAU~\cite{shende2006tau} are two such examples. Additionally, literature also includes targeted performance analysis tools for automatic identification of inefficient synchronization regions and memory transfers~\cite{benjamin2019diogenes}, optimization strategies for improving instruction pipelining, and memory access performance~\cite{zhang20122a}, and reducing P2P communication across multi GPUs~\cite{chen2018towards}. Bateni et al. ~\cite{bateni2020cooptimizing} focus on improving the application execution time by dynamically selecting memory management policies to relieve pressure on system memory. In contrast, this research targets identifying application-device interactions to optimize both device utilization and application performance. Due to the significant role of data movement in GPU computing, plenty of work has been done on managing CPU-GPU communication. A plethora of research exists on tuning data movement for performance improvement, e.g., exploration of the trade-off in data transfer overhead on discrete vs. integrated GPU systems~\cite{zhou2018s}, dynamic work partitioning to mitigate performance variability on GPUs~\cite{boyer2013load}, partitioning work between CPU and GPU to accelerate certain tasks~\cite{delorme2013parallel} and reducing the pending time of high-priority tasks through preemptive interception~\cite{zhou2015gpes}.

As integrated GPU systems are becoming important, especially in autonomous computing, the focus has shifted to unifying CPU-GPU memory to reduce programming burden and increase system performance~\cite{ganguly2019interplay}. Analysis of data access patterns and data transfer performance in CUDA under the UMA mechanism has also been proposed~\cite{landaverde2014investigation}. Hestness et al.~\cite{hestness2015gpu} analyze opportunities to optimize computing and cache efficiency on integrated GPU architectures. Compared to others, our work presents a systematic approach to understanding how GPU applications utilize different hardware components and how well these interactions explain application performance, device utilization, or both. Insights from such an analysis can then be used to systematically explore code transformations to improve a performance metric of choice (e.g., execution time, compute resource utilization). Given that the state-of-the-art focuses squarely on application performance, our approach of targeting both device utilization and performance is novel.

\section{Conclusion}
In this paper, we present a composite performance metric for characterizing application performance to identify application-device interactions that affect it. We also propose a comparative analysis technique that combined with the proposed metric, can be applied to characterize code transformation techniques to discern application-device interactions induced by code transformations and their impact on the multi-objective performance metric. The objective of this work is to develop a systematic approach for identifying code transformations to improve several performance metrics and is done in two steps--first, to characterize an application in regards to a performance metric of choice, and then guided by the resource utilization behavior of the application, select transformations that reduce expensive application-device interactions. 
Specifically, in this work, we propose a metric to quantify both application performance and device utilization. 
This paper studies the application-device interactions of several GPU applications and computes resource-wise importance, understand how resource usage behaviors impact application performance and device utilization, and finally connects the two by proposing optimizations based on the hardware resource usage behaviors and demonstrating improvements in both performance and device utilization. 
We then apply these code transformations, which improve the execution time up to 29.6\% (27.4\% on average) and device utilization up to 5.4\% (2\% on average). The methodologies and performance data will be made publicly available on Github. This research identifies the critical application-device interactions for jointly improving the application performance and device utilization on a single GPU, thus building a foundation for future multi-GPU studies.

\section{Acknowledgement}
This material is based upon work supported by the U.S. Department of Energy, Office of Science under Award Number DE-SC0022843.

\begingroup
    \bibliographystyle{IEEEtran}
    \bibliography{main}
\endgroup

\begin{IEEEbiography}[{\includegraphics[width=1in,height=1.25in,clip,keepaspectratio]{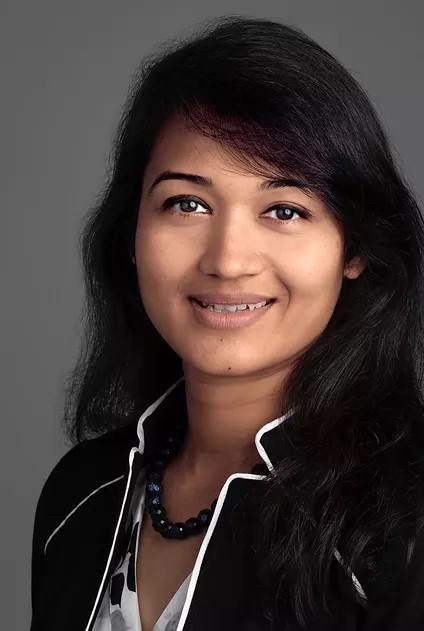}}]{Dr. Tanzima Islam}
Dr. Tanzima Islam is an assistant professor in the Department of Computer Science at Texas State University (TxState). Dr. Islam earned her Ph.D. in Computer Engineering from Purdue University and was a postdoctoral scholar at Lawrence Livermore National Laboratory (LLNL). Her research develops software tools and data-driven analysis techniques to automatically identify performance problems of scientific applications running on the High-Performance Computing (HPC) systems and mitigate them. Dr. Islam is a DOE SRP fellow and has co-authored over 30 peer-reviewed publications at top-tier conferences and journals in the area of HPC performance. She has received numerous awards for the impact of her research, including a 2022 DOE Early Career Award, 2019 R\&D 100 award, Science and Technology award from LLNL, and the College of Science and Engineering’s Excellence in Scholarly Activities at TxState. Dr. Islam’s research has been funded by national labs and the industry such as DOE, LLNL, AMD.\end{IEEEbiography}

\begin{IEEEbiographynophoto}{Dr. Aniruddha Marathe}
Dr. Aniruddha Marathe is a Computer Scientist at the Center for Applied Scientific Computing (CASC) at the Lawrence Livermore National Laboratory (LLNL). His research focuses on developing performance-optimizing run-time algorithms for HPC applications on resource-constrained clusters, power-aware computing and feedback-driven performance tuning. Before joining LLNL, he was a Postdoctoral Research associate at the Department of Computer Science at The University of Arizona. He received his Doctorate in Computer Science from the Department of Computer Science at The University of Arizona in 2014. Dr. Marathe has co-authored over 20 peer-reviewed publications at top-tier conferences and journals in the area of system performance tuning. He is currently the lead contributor to the power-aware computing runtime software in the Exascale Computing Project (ECP) initiative, a founding member of the bi-annual PowerStack seminar and has served as the co-PI for the NASA SBIR grant on feedback-driven performance optimization. He is a member of ACM. 
\end{IEEEbiographynophoto}

\begin{IEEEbiography}[{\includegraphics[width=1in,height=1.25in,clip,keepaspectratio]{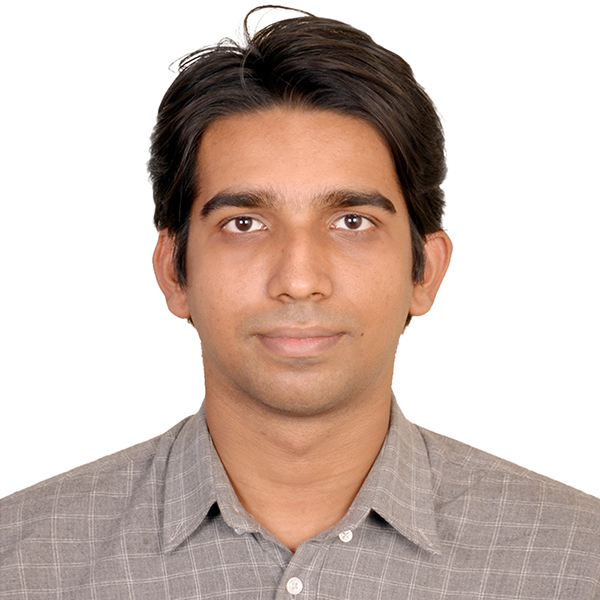}}]{Mohammad Zaeed}
Mohammad Zaeed is a Ph.D. student in the Department of Computer Science at Texas State University. He earned his Bachelor's in Software Engineering from the Institute of Information Technology at the University of Dhaka and was a lecturer at the Bangladesh University of Business and Technology. He is currently working on analyzing the performance of high-scale applications. 
\end{IEEEbiography}
\end{document}